\documentclass{iau}
\usepackage{graphicx}
\usepackage{subfigure}
\usepackage{float}

\title[JD 11.~~Observational Study of Large Amplitude Longitudinal Oscillations in a Solar Filament ] 
{Observational Study of Large Amplitude Longitudinal
Oscillations in a Solar Filament }
\vspace{-1cm}
\author[K. Knizhnik, M. Luna, K. Muglach, H. Gilbert, T. Kucera, J. Karpen]   
{Kalman Knizhnik$^{1,2},$
 Manuel Luna$^3$,
 Karin Muglach$^{2,4}$\\
 Holly Gilbert$^2$,
 Therese Kucera$^2$,
 Judith Karpen$^2$}

\affiliation{$^1$Department of Physics and Astronomy \\ 
The Johns Hopkins University, Baltimore, MD 21218\\
email: {kalman.knizhnik@nasa.gov} \\[\affilskip]
$^2$NASA/GSFC, Greenbelt, MD 20771, USA\\
$^3$ Instituto de Astrof{\'i}sica de Canarias,
E-38200 La Laguna, Tenerife, Spain \\ 
$^4$ ARTEP, Inc., Maryland, USA}

\pubyear{2014}
\volume{300}  
\pagerange{xxx}
\jname{Nature of Prominences and their role in Space Weather}
\editors{B. Schmieder, JM. Malherbe \& S. Wu, eds.}
\begin{document}

\maketitle
\vspace{-0.2cm}
\begin{abstract}
On 20 August 2010 an energetic disturbance triggered damped large-amplitude longitudinal (LAL) oscillations in almost an entire filament. In the present work we analyze this periodic motion in the filament to characterize the damping and restoring mechanism of the oscillation.  Our method involves placing slits along the axis of the filament at different angles with respect to the spine of the filament, finding the angle at which the oscillation is clearest, and fitting the resulting oscillation pattern to decaying sinusoidal and Bessel functions. These functions represent the equations of motion of a pendulum damped by mass accretion. With this method we determine the period and the decaying time of the oscillation. Our preliminary results support the theory presented by Luna and Karpen (2012) that the restoring force of LAL oscillations is solar gravity in the tubes where the threads oscillate, and the damping mechanism is the ongoing accumulation of mass onto the oscillating threads. Following an earlier paper, we have determined the magnitude and radius of curvature of the dipped magnetic flux tubes hosting a thread along the filament, as well as the mass accretion rate of the filament threads, via the fitted parameters.
\vspace{-0.2cm}
\end{abstract}
\firstsection 
\vspace{-0.35cm}
\section{Procedure}
LAL oscillations consist of periodic motions of the prominence threads along the magnetic field that are disturbed by a small energetic event close to the filament (see Luna et al. paper in this volume). Luna and Karpen (2012) argue that prominence oscillations can be modeled as a damped oscillating pendulum, whose equation of motion satisfies a zeroth-order Bessel function. In their model, a nearby trigger event causes quasi-stationary preexisting prominence threads sitting in the dips of the magnetic structure to oscillate back and forth, with the restoring force being the projected gravity in the tubes where the threads oscillate (e.g. Luna et al. (2012)). In this paper, we report preliminary results of comparisons of observations of prominence oscillations with the model presented by Luna and Karpen (2012). More details will be available in the forthcoming paper by Luna et al. (2013). \par
\begin{figure}[h]\label{figures}
\begin{center}
\hbox{
\hspace{-1cm}
	\includegraphics[height=5.8cm,width=5.8cm]{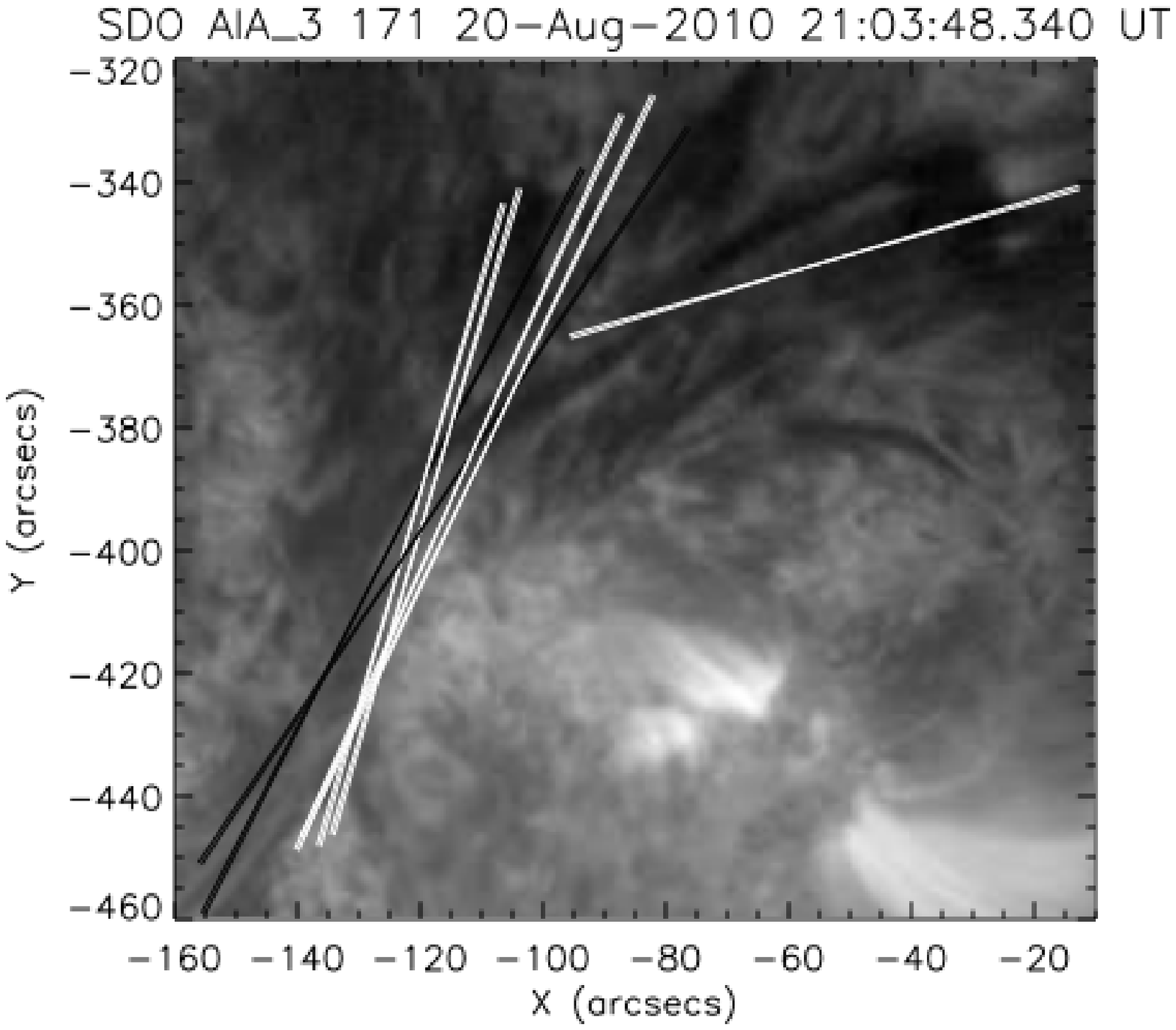}
	 \includegraphics[height=5.7cm,width=8.8cm]{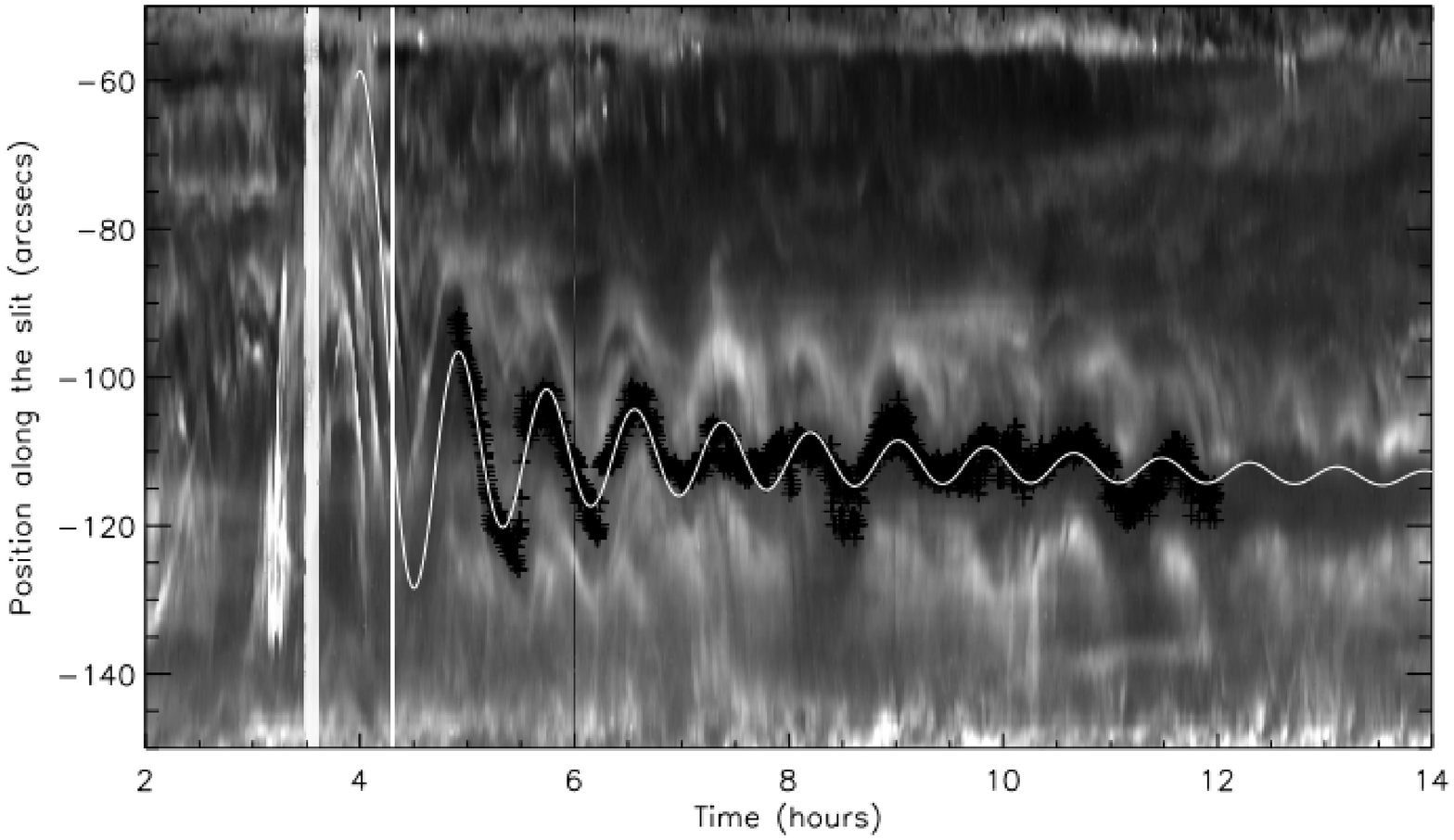}
	}
\caption{\emph{Left}: Filament seen in AIA 171 with best slits overlaid. \emph{Right}: An intensity distance-time slit, showing an oscillation with the Bessel fit (white curve) to equation (2.1) in Luna et al. (this volume). The sinusoidal fit was not as good as the Bessel fit and is not shown.}	
\end{center}
\end{figure}

In this analysis, we place slits along the filament spine and measure the intensity along each slit as a function of time. Fig. \ref{figures} (left) shows the filament in the AIA 171\AA\ filter with the slits overlaid. Each slit is then rotated in increments of 0.5$^\circ$ from 0$^\circ$ to 60$^\circ$ with respect to the filament spine. We select the best slit according to the following criteria: 
(a) continuity of oscillations,
(b) amplitude of the oscillation is maximized,
(c) clear transition from dark to bright regions, 
(d) maximum number of cycles.
\vspace{-0.5cm}

\vspace{0.5cm}
The oscillation for a representative slit is shown in Figure \ref{figures} (right), which corresponds to the grey slit in Figure \ref{figures} (left). We identify the position of the center of mass of the thread by finding the intensity minimum along the slit, indicated by black crosses in Figure \ref{figures} (right). These points are then fit to equation (2.1) of Luna et al. (this volume), and the resulting fit is shown in white. 

\vspace{-0.6cm}
\section{Results}\label{results}
Fitting our data to equation (2.1) of Luna et al. (this volume) yields values of $\chi^2$ ranging between 1-13. Using equation (2.2) of Luna et al. (this volume), we find the average radius of curvature of the magnetic field dips that support the oscillating threads. We find it to be approximately 60 Mm. We also calculate a threshold value for the field itself that would allow it to support the observed threads. Using equation (3.1) of Luna et al. (this volume), we find an average magnetic field of $\sim20 \;G$, assuming a typical filament number density of $10^{11}$ cm$^{-3}$,  in good agreement with measurements (e.g. Mackay et al. 2010). On average, the oscillations form an angle of $\sim25 ^o$ with respect to the filament spine, and have a period of $\sim 0.8$ hours. To explain the very strong damping mass must accrete onto the threads at a rate of about 60 $\times10^6$ kg/hr. 
\vspace{-0.55cm}
\section{Conclusions}
We conclude that the observed oscillations are along the magnetic field, which forms an angle of $\sim$$25^o$ with respect to the filament spine (Tandberg-Hanssen \& Anzer, 1970). 
We find that both the curvature and the magnitude of the magnetic field are approximately uniform on different threads. Both the Bessel and sinusoidal functions are well fitted, 
indicating that mass accretion is a likely damping mechanism of LAL oscillations, and that the restoring force is the projected gravity in the dips where the threads oscillate. 
The mass accretion rate agrees with the theoretical value (Karpen et al., 2006, Luna, Karpen, \& DeVore, 2012).


\vspace{-0.67cm}
\begin{thebibliography}{}
\vspace{-0.09cm}
\bibitem[Karpen \etal\ (2006)]{Karpen2006}
Karpen, J. T., Antiochos, S. K., Klimchuk, J. A. 2006, ApJ, 637, 531

\bibitem[Luna \etal\ (2012a)]{luna2012a}
Luna, M., Karpen, J.~T., \& Devore, C.~R. 2012a, \textit{ApJ}, 746,  30
\bibitem[Luna \& Karpen (2012)]{luna2012b}
Luna, M., \& Karpen, J. 2012, \textit{ApJ}, 750, L1

\bibitem[Luna \etal\ (2013)]{luna2013thisvol} Luna, M., Knizhnik, K., Muglach, K., Gilbert, H, Kucera, T. \& Karpen, J., \textit{this volume}, 2014
\bibitem[Luna \etal\ (2013)]{luna2013} Luna, M., Knizhnik, K., Muglach, K., Gilbert, H, Kucera, T. \& Karpen, J., \textit{ApJ}, 2013, \textit{in prep.}

\bibitem[Mackay \etal\ (2010)]{mackay2010}
{Mackay}, D., {Karpen}, J., {Ballester}, J., {Schmieder}, B.,
  {Aulanier}, G. 2010, \textit{Sp. Sci. Rev.}, 151, 333


\bibitem[Tandberg-Hanssen \& Anzer (1970)]{TH1970}
{Tandberg-Hanssen}, E. and {Anzer}, U. 1970, \textit{Solar Physics} 15, 158T




\end{thebibliography}
\end{document}